\documentclass[aps,pra,twocolumn,tightenlines]{revtex4-1}
\usepackage{amsmath,epsfig,amssymb}
\usepackage{bbm}
\usepackage{times}
\usepackage{color}
\usepackage{amsthm}
\usepackage{color}
\usepackage{rotating}
\usepackage[caption=false]{subfig}
\usepackage{algorithm}
\usepackage{algpseudocode}

\newcommand{\todo}[1]{}
\renewcommand{\todo}[1]{{\color{red} TODO: {#1}}}
\newcommand{\question}[1]{}
\renewcommand{\question}[1]{{\color{red} QUESTION: {#1}}}
\renewcommand\Re{\operatorname{Re}}

\DeclareMathOperator*{\argmax}{arg\,max}
\DeclareMathOperator{\sign}{sign}
\DeclareMathOperator{\Tr}{Tr}

\begin{document}

\title{Exponential Families for Bayesian Quantum Process Tomography}
\author{Kevin Schultz}
\email{kevin.schultz@jhuapl.edu}
\affiliation{
 Johns Hopkins University Applied Physics Laboratory\\
 11100 Johns Hopkins Road, Laurel, MD, 20723, USA
}

\date{\today}


\begin{abstract}

A Bayesian approach to  quantum process tomography has yet to be fully developed
due to the lack of appropriate probability  distributions on the space of
quantum channels. Here, by associating the Choi matrix form of a completely
positive, trace preserving (CPTP) map with a particular space of matrices with
orthonormal columns, called a Stiefel manifold, we present two parametric
probability distributions on the space of CPTP maps that enable Bayesian
analysis of process tomography. The first is a probability distribution that
has an average Choi matrix as a sufficient statistic. The second is a
distribution resulting from binomial likelihood data that enables a simple
connection to data gathered through process tomography experiments.  To our
knowledge these are the first examples of continuous, non-unitary random CPTP
maps, that capture meaningful prior information for use in  Bayesian
estimation.   We show how these distributions can be used for point estimation
using either maximum a posteriori estimates or expected a posteriori estimates,
as well as full Bayesian tomography resulting in posterior credibility
intervals. This approach will enable the full power of Bayesian analysis in all
forms of quantum characterization, verification, and validation.

\end{abstract}

\maketitle

\section{Introduction}
In the quest to develop quantum technologies, precise characterization of
quantum systems is critical to understanding and mitigating noise and
imperfections. To date, nearly all quantum characterization techniques such as
randomized benchmarking (RB) \cite{knill2008randomized}, quantum state and process
tomography (see e.g. Refs.  \cite{jones1994fundamental,chuang1997prescription,
poyatos1997complete,Merkel2013,blume2016certifying}), gate-set tomography (GST)
\cite{Merkel2013,blume2016certifying}, or even more advanced techniques such as
quantum noise spectroscopy \cite{yuge2011measurement,alvarez2011measuring} tend
to primarily rely on a frequentist approach where a large number of
experimental data-points are taken in order to estimate the average of some
parameter or parameters of interest.  As error-rates continue to march lower,
having precise knowledge of uncertainties in the estimated parameters is key to
improving devices below the fault-tolerant threshold needed for large-scale
quantum computing.

To get a sense of the estimation accuracy and to reduce experimental costs,
Bayesian-based approaches for RB
\cite{granade2015accelerated,hincks2018bayesian} process tomography
\cite{granade2016practical,teo2018bayesian}, and noise spectroscopy
\cite{ferrie2018bayesian} have been proposed. Despite this intense interest, a
fully Bayesian approach to quantum process tomography with analytic prior and
posterior distributions has yet to be developed, with the current approaches
relying on importance sampling and simple proposal distributions
\cite{granade2016practical}.

A key requirement for Bayesian estimation is having a probability distribution
or distributions defined on the relevant space of interest. Furthermore it is
generally desired that these distributions can be parameterized in some manner
that allows for the capture of meaningful prior information.  Typically, this
amounts to some form of location and scale parameters. As an example the normal
distribution is defined on the space of real numbers and is fully parameterized
by a mean and variance.  In the context of quantum process tomography, one
particular sample space of interest is the space of completely positive and
trace preserving (CPTP) maps (see e.g. \cite{nielsen2010quantum}). Therefore, a
probability distribution on the space of CPTP maps, preferably with a simple
parameterization, could enable fully Bayesian quantum process tomography.

Perhaps the simplest and most familiar example of a random quantum operation
are Pauli error channels, commonly used in quantum error correction simulations
(see e.g.
\cite{aaronson2004improved,knill2005quantum,gutierrez2013approximation}).  On
one hand, the average of these distributions is well defined and simple to
calculate, allowing them to be used in the analysis of noisy quantum circuits.
Their application in Bayesian setting is limited, however.  These distributions
are fundamentally discrete, and thus are not absolutely continuous with respect
to the entire space of CPTP maps and as such will necessarily produce estimates
that are Pauli channels and cannot capture error effects such as coherent
rotations and non-unital errors that will show up in any non-idealized
tomographic procedure.
A slight generalization of this concept is discrete mixtures over a finite set
of CPTP maps \cite{audenaert2008random}, which has the same issues as Pauli
error channels in the Bayesian context.

More exotic examples of probability distributions on the space of CPTP maps
come from the field of random matrix theory. Projections of Haar-random
unitaries from a unitary group of higher dimension has been used to define a
distribution on the space of CPTP maps that is absolutely continuous on the
space of CPTP maps~\cite{bruzda2009random}. This essentially provides a uniform
distribution of CPTP maps with a given Kraus rank.
Ref.~\cite{collins2016random} further provides a useful review on a number of
results on random CPTP maps derived from random unitary operations, along with
a number of theoretical results on their composition and effect on quantum
states. The downside to these approaches is that there is no particularly
useful parameterization in terms of a location and scale parameter. In fact the
average of these distributions is the maximally mixed channel. This is not
particularly useful as a prior for a high-fidelity quantum channel, which is
what one requires when characterization quantum devices.  That said, the
distribution of \cite{bruzda2009random} has been used as a proposal
distribution for process tomography in \cite{pogorelov2017experimental} using
similar importance sampling techniques as in \cite{granade2016practical}.  The
efficiency of this approach would be greatly enhanced by a proposal
distribution that can capture meaningful prior information, such as the distributions
introduced below.

In contrast to the above approaches, here we demonstrate two approaches to
generating probability distributions relevant to quantum process tomography.
The first is a family of probability distributions of random CPTP maps that are
absolutely continuous and completely characterized by their first moment, in
this case, an average Choi matrix. The second is an alternative formulation
where these distributions can be characterized as maximum entropy distributions
with a given average value. Our approach makes use of the theory of exponential
families on the manifold of matrices with orthonormal columns, known as Stiefel
manifolds \cite{james1976topology}.

Our first approach is natural to consider in the context of noisy quantum
systems.  On one hand, the evolution of an open quantum system can often be
treated as a noisy evolution, through the stochastic Liouville equation
\cite{kubo1963stochastic}.  This type of description applies to some of the
most common types of qubits and their predominant decoherence mechanisms (see
e.g. Refs. \cite{Kubo1957, Schulten1978, ernst1987principles, Schneider1998,
Grigorescu1998, abergel2003, Cheng2004, Wilhelm2007}).  On the other hand,
especially in the context of quantum gates or circuits, we often speak of just
an average error channel (or simply ``the'' error channel) for a given quantum
operation, and do not consider it as either a random variable or stochastic
process.  In this context, quantum process tomography is essentially computing
estimates of the average quantum channel. This motivates the desire for a
parametric probability distribution for which the average map, as estimated
from tomography or simulation, is a sufficient statistic. This approach
provides such a parameterization.

The second approach is derived from the likelihood of a process tomographic
experiment using the same manifold of matrices with orthonormal columns. This
in turn allows for construction of a second class of prior based on the
conjugate prior for binomial data.  In effect, this prior uses synthetic
``pseudo-experiments'' to capture prior information about the distribution of
CPTP maps.  Since this distribution is derived from the likelihood of binomial
data, it is also the posterior distribution for binomial process tomographic
data.  Using a sampling scheme adapted from \cite{hoff2009simulation}, we show
how these two priors can be combined to perform Bayesian process tomography for
both point estimation and for generating full posterior distributions. This
allows one to efficiently include prior information from techniques such as RB
into process tomography experiments.

In the following sections, we relate known properties of CPTP maps to Stiefel
manifolds and show how this representation is compatible with process
tomography. Next we review definitions from classical statistics and introduce
the concept of an \textit{exponential family} of probability distributions, and
derive two families on the Stiefel manifold for the purposes of Bayesian
process tomography.  Following this, we show how these distributions can be
used to generate Bayesian estimates for process tomography, and compare the
results to traditional maximum likelihood methods.

\section{CPTP Maps and Stiefel Manifolds}
In quantum information, a quantum state can be represented by a density
operator $\rho$, where $\rho\in\mathbb{C}^{N\times N}$ is a positive
semi-definite, Hermitian matrix with $\Tr(\rho) =1$.
Quantum operations are then completely positive, trace-preserving (CPTP) maps
\cite{nielsen2010quantum}.  In this work, we will make the additional assumption that
the quantum maps of interest map to density operators of the same dimension as
the input dimension, but this can be generalized. CPTP maps can be represented
 by the Choi matrix form $\Lambda$, which can be derived from a
Liouvillian superoperator $\mathcal{L}$ via a coordinate shuffling involution
operation \cite{bruzda2009random,wood2015tensor}.  The relevant properties of
$\Lambda$ that we will consider here are 1) the CP property implies $\Lambda$
is Hermitian and postive-semidefinite, and 2) the TP property implies that
$\Tr_B\Lambda=I_N$, where $\Tr_B$ denotes the partial trace over the second
subsystem when $\Lambda$ is viewed as an operator in the tensor product space
of two $N\times N$ spaces.
Since $\Lambda$ is Hermitian and positive semidefinite, there exists a matrix
$S$ such that $\Lambda=S^\dagger S$, i.e., a square root factorization.
Note that this factorization is not unique, indeed $US$ for any unitary $U$ of
appropriate dimension will result in an identical Choi matrix as $S$.  
Furthermore, even the dimension of $S$ is not unique as the rank $K$ of
$\Lambda$ (i.e., the Kraus rank) implies that $S$ can be expressed as a
$K\times N^2$ complex matrix, but there exist $M\times N^2$-dimensional matrix
factorizations for all $M\geq K$.
Regardless of the specific choice of $S$, the coordinates of $\Lambda$ can be
expressed as inner products of the columns of $S$, through
$\Lambda_{ij}=\langle S_i,S_j\rangle=S_i^\dagger S_j$.  Consider next the
$NM\times N$ complex matrix derived from an $M\times N$ square root factorization $S$
\begin{equation}\label{eq:decomp}
    \xi=\begin{bmatrix} S_1 & S_{N+1} & \dots & S_{N(N-1)+1}\\
            S_2 & S_{N+2} & \dots & S_{N(N-1)+2}\\ 
            \vdots & \vdots &\ddots & \vdots\\
            S_{N} & S_{2N} & \dots & S_{N^2} \end{bmatrix}\,.
\end{equation}
First, note 
\begin{equation}
    \Tr_B(\Lambda)=I_N\implies \sum_{i=1}^{N}\Lambda_{kN+i,kN+i}=1\,,
\end{equation}
for $k=0,\dots,N-1$.  Thus, 
\begin{equation}
    ||\xi_{k+1}||_2^2=\sum_{i=1}^{N}S_{kN+i}^\dagger S_{kN+i}=1\,,
\end{equation}
so the columns of $\xi$ are unit vectors.  Second,
$$\Tr_B(\Lambda)=I_N\implies \sum_{i=1}^N S_{jN+i}^\dagger S_{kN+i}=0,$$ for
$j,k=0,\dots,N-1$ and $j\neq k$, so the columns of $\xi$ are in fact
orthonormal with $\xi^\dagger\xi=I_N$.  The space of $m\times n$ (complex)
matrices  ($m\geq n$) with orthonormal columns is a Stiefel manifold
\cite{james1976topology}, and is denoted $V_{n}(\mathbb{C}^m)$.  We will show
below how a certain probability distribution on Stiefel manifolds corresponds
to a natural distribution of random CPTP maps, and as such we restrict our
attention to the Stiefel manifolds $V_{N}(\mathbb{C}^{kN})$, where
$k=1,\dots,N^2$ depending on the Kraus rank of the random CPTP maps we are
working with.

This is not the first time that a CPTP map has been associated with elements of
a Stiefel manifold, it has been noted elsewhere (see e.g.,
\cite{bruzda2009random, pechen2014incoherent,pogorelov2017experimental}) that
column stacking the Kraus operators in the Kraus form of a CPTP map can be
associated with an equivalence class of unitary matrices with the same $N$
columns (see \cite{edelman1998geometry} for a characterization of Stiefel
manifolds in terms of these equivalence classes).  Using the eigendecomposition
of the Choi matrix, one can construct Kraus operators \cite{wood2015tensor} and
these can in turn be column stacked and identified with an element of a Stiefel
manifold.  Careful index tracking reveals that the Stiefel representation $\xi$
can be mapped to stacked Kraus operators through a row permutation.  These
representations of CPTP maps are known collectively as Stinespring
representations \cite{stinespring1955positive,wood2015tensor}, and were used in
Ref.~\cite{bruzda2009random} as an alternative derivation for the generation of
their random CPTP map distribution.

\subsection{The Born Rule and Stiefel Manifolds}

For an orthogonal set of projective measurement operators $\{F_i\}$ and a density operator
$\rho$, $p_i$, the probability of recording measurement outcome $i$ is
$p_i=\Tr(F_i\rho)=\langle\langle F_i||\rho\rangle\rangle$ where
$|\cdot\rangle\rangle$ denotes the column-stacking vectorization operator, and
$\langle\langle \cdot|$ its conjugate transpose.  For a given quantum process
with Liouvillian $\mathcal{L}$ and initial density operator $\rho$ we have
$p_i=\langle\langle
F_i|\mathcal{L}|\rho\rangle\rangle=\Tr(|\rho\rangle\rangle\langle\langle F_i|\mathcal{L})=\Tr((|F_i\rangle\rangle\langle\langle
p|)^\dagger\mathcal{L}|)$.

Next, let $\mathfrak{P}$ be the coordinate shuffling involution that maps
Liouvillian superoperators to Choi matrices
\cite{bruzda2009random,wood2015tensor}, so that
$\Lambda=\mathfrak{P}(\mathcal{L})$. Since $\mathfrak{P}$ is both coordinate
shuffling and an involution (i.e.,
$\mathfrak{P}(\mathfrak{P}(\mathcal{L}))=\mathcal{L}$) then $\mathfrak{P}$ is a
permutation and thus a unitary ``super-duper operator,'' and as such preserves
inner products between the two spaces.  This implies that $p_i$ can be
expressed as $p_i=\Tr(\mathfrak{P}(|F_i\rangle\rangle\langle\langle
\rho|)^\dagger\mathfrak{P}(\mathcal{L}))=\Tr(\mathfrak{P}(|F_i\rangle\rangle\langle\langle\rho|)^\dagger\Lambda)$.

Next, let $S$ be a square root of $\Lambda$ and $\xi$ be a corresponding
Stiefel manifold representation  with Kraus rank $K$.  Then, for an arbitrary
matrix $N^2\times N^2$ complex matrix $\Theta$, we have  
\begin{equation}\begin{aligned}\label{eq:born_stiefel1}
    \Tr\left(\Theta^\dagger \Lambda(\xi)\right) & = \Tr\left(\Theta^\dagger
    S^\dagger S\right)\\ &=\sum_{i,j=1}^{N^2}\Theta_{ij}^*S_i^\dagger S_j\\
    &=\langle\langle S|\left(\Theta^\dagger\otimes
    I_{K}\right)|S\rangle\rangle\\ &=\langle\langle\xi|\left(\Theta^\dagger
    \otimes I_{K}\right)|\xi\rangle\rangle\\
\end{aligned}\end{equation}
and thus for a given $F_i$ and $\rho$, we can express the output probability
$p_i$ as 
\begin{equation}\label{eq:born_stiefel2}
p_i = \langle\langle\xi|\left(\mathfrak{P}(|F_i\rangle\rangle\langle\langle
    \rho|)^\dagger\otimes I_{K}\right)|\xi\rangle\rangle\,.
\end{equation}
Equation~\eqref{eq:born_stiefel1} can also be expressed in terms of the $N^2$
$KN$-dimensional columns of $\xi$ by
\begin{equation}\begin{aligned}\label{eq:born_stiefel3}
    \langle\langle\xi|(\Theta^\dagger\otimes I_K)|\xi\rangle\rangle&=
    \sum_{i,j=1}^{N}\xi_i^\dagger (\Theta^\dagger\otimes I_K)_{[i,j]}\xi_j
\end{aligned}\end{equation}
where $_{[i,j]}$ indexes over $KN\times KN$ subblocks, and analogously for
Eq.~\eqref{eq:born_stiefel2}.

At first glance, the expressions in
Eqs.~\eqref{eq:born_stiefel1}-\eqref{eq:born_stiefel3} may appear cumbersome
and complicated.  In the following sections, we will show that since the
Stiefel manifold representation $\xi$ naturally encodes the constraints of a
CPTP map, Eqs.~\eqref{eq:born_stiefel1}-\eqref{eq:born_stiefel3} can be used
not only to define natural prior and posterior distributions on Stiefel manifolds, but
are also essential to efficient sampling routines of these distributions.

\subsection{Basic Process Tomography}\label{sec:basic_pt}

Process tomography refers to the estimation of a quantum channel given a set of
experimental setups and outcomes \cite{chuang1997prescription}. Here we will
focus on a basic setup with error-free arbitrary state preparation and measurement.
This allows us to demonstrate the ideas in future sections, but in principle these
restrictions could be removed and the techniques adapted to more advanced forms
of process tomography such as \cite{Merkel2013,blume2016certifying}.
In an ideal experiment with perfect state preparation and projective
measurements, a given measurement is a Bernoulli trial, and repeating the
experiment results in a binomial distribution.  If we perform $m$ different
state preparation $\rho_i$ and measurement combinations $F_i$, accumulating
$x_i$ counts in $n_i$ trials, the resulting joint binomial probability
distribution is
\begin{align}\label{eq:likelihood}
    p(x_1,\dots,x_m & | n_1,\dots,n_m,\Lambda) = \\ \nonumber &
    \prod_{i=1}^m\binom{n_i}{x_i}\Tr(A_i^\dagger\Lambda)^{x_i}
    (1-\Tr(A_i^\dagger\Lambda))^{n_i-x_i},
\end{align}
where $A_i=\mathfrak{P}(|F_i\rangle\rangle\langle\langle \rho_i|)$.
%
%
For notational convenience, we will use bold-faced $\mathbf{x}$ and
$\mathbf{n}$ to denote the $m$-dimensional count and trial vector, respectively
so that
\begin{equation}
    p(\mathbf{x}|\mathbf{n},\Lambda)\triangleq(x_1,\dots,x_m|n_1,\dots,n_m,\Lambda)\,.
\end{equation}

Given the data $\mathbf{x}$, $\mathbf{n}$, and experimental setups $A_i$, a
common method for producing an estimate $\hat{\Lambda}$ of the unknown channel
$\Lambda$, is to perform a least squares fit of the system of equations implied
by $\langle\langle A_I||\Lambda\rangle\rangle=x_i/n_i$with appropriate
constraints to ensure that $\Lambda$ is CPTP.
Alternatively, maximum likelihood estimation of the Choi matrix in process
tomography is formally defined as
\begin{equation}\label{eq:MLE}
    \hat{\Lambda}_{MLE} = \argmax_{\Lambda \textrm{
        CPTP}}\,\,p(\mathbf{x}|\mathbf{n}, \Lambda)
\end{equation}
where $\Lambda$ is drawn from the space of Choi matrices corresponding to  CPTP
maps.  Typically, maximum likelihood estimation (MLE) is performed using the
equivalent optimization of the log of the likelihood function
Eq.~\eqref{eq:likelihood} by
\begin{align}\label{eq:logMLE}
    \hat{\Lambda}_{MLE} & = \argmax_{\Lambda\textrm{ CPTP}}\,\, \sum_{i=1}^m
    x_i\log \Tr(A_i^\dagger\Lambda) \\ \nonumber &
    +(n_i-x_i)\log(1-\Tr(A_i^\dagger\Lambda)) \\ \nonumber & +\textrm{ terms
    constant in }\Lambda.
\end{align}
More complex tomographic procedures, such as gate set tomography
\cite{Merkel2013, blume2016certifying}, can use a similar objective function,
except it is jointly optimized over multiple CPTP maps and the
$\Tr(A_i^\dagger\Lambda)$ are replaced with higher order terms in the gates to
be estimated.  The least squares technique described above can also be cast
in terms of maximum likelihood by using the Gaussian approximation to the
binomial distribution.


\section{Exponential Families of CPTP Maps}
In classical statistics, one is often concerned with the estimation of some
parameter $\vartheta$ given sample data $\{x_i\}$ using some family of
parameterized probability distributions $p(x;\vartheta)$. In particular, it is
desirable to select statistical models $p(x;\vartheta)$ where sample averages
of some function $T(x)$ contain all of the information needed for the maximum
likelihood estimation of $\vartheta$ that can be derived from a dataset $\{x_i\}$.
In this context, $T(x)$ is referred to as a \textit{sufficient statistic}
\cite{koopman1936distributions,pitman1936sufficient}. A familiar example of
this concept is the estimation of the parameters of a normal
random variable through the sample averages of the mean and variance from data.

An alternative method for the selection of statistical models uses the
principle of maximum entropy \cite{cover2012elements}, where the functional
form of the probability distribution is selected based on maximizing the
entropy functional $E[\log(p(x))]$ over the space of all probability
distributions that satisfy $E[T(x)]=\eta$ relative to some base measure,
defining a distribution $p(x;\eta)$.  Again, an example of this is the normal
distribution, which maximizes the entropy over all probability distributions
(with support on the entirety of $\mathbb{R}^n$) with a given mean and
variance, relative to the Lebesgue measure on $\mathbb{R}^n$.

More generally, it turns out that under a very broad set of conditions, these
two methods of defining probability distributions lead to identical
distributions.  There exists a 1-1 differentiable mapping between the
parameters $\vartheta$ and $\eta$, and the study of these dual coordinates is
known as information geometry \cite{amari2007methods}, which generalizes a wide
range of properties encountered in many familiar probability distributions.
The generalization of these concepts leads to the notion of exponential
families, which are parametric families of probability distributions with the
following form:
\begin{equation}
\label{eq:exp_family} p(x;\vartheta) = \exp\left(\langle\vartheta, T(x)\rangle
    -\psi(\vartheta)+\kappa(x)\right)
\end{equation}
where $\vartheta$ are called the natural parameters that enforce the above
decoupling between $x$ and the parameters, $T(x)$ are the sufficient
statistics, $\psi(\vartheta)$ is the log-normalizer which forces
$p(x;\vartheta)$ to integrate to 1, and $\kappa(x)$ is the carrier measure
which defines the support of $p(x;\vartheta)$ in the full space of $x$.  It is
easy to check that maximum (log)-likelihood estimation of $\vartheta$ depends
only sample averages of $T(x)$, as desired. A more complicated argument using
Lagrange multipliers can be used to show that this distribution maximizes the
entropy subject to constraints on the sufficient statistics. Furthermore,
exponential families are essentially the only distributions that satisfy this
property \cite{koopman1936distributions,pitman1936sufficient}.  As implied
above, the normal distribution is an example of an exponential family
($\vartheta=(\frac{\mu}{\sigma^2},\frac{-1}{2\sigma^2})^\top$,
$T(x)=(\mu,\mu^2+\sigma^2)^\top$, and
$\psi(\vartheta)=\frac{\vartheta_1^2}{4\vartheta_2}+\frac{1}{2}\log(-\pi/\vartheta_2)$),
and another particularly relevant example is the binomial distribution
($\vartheta=\log(p/(1-p))$, $T(x)=np$, and
$\psi(\vartheta)=n\log(1+\exp(\vartheta))-\log(n!)$).
%

\subsection{Choi Matrices as Sufficient Statistics}\label{sec:suff}
To define a probability distribution on the space of CPTP maps for which the
average Choi matrix is a sufficient statistic (alternatively, one for which
entropy is maximized given an average Choi matrix), we arrive at an exponential
family of the form:
\begin{equation}\begin{aligned}\label{eq:suff1}
    p(\Lambda;\Theta)=\exp\left(\Tr(\Theta^\dagger \Lambda) -
    \psi(\Theta)+\kappa(\Lambda)\right)\,,
\end{aligned}\end{equation}
where $\Theta$ denotes the (matrix) natural parameter to the corresponding
sufficient statistic $T(\Lambda)=\Lambda$.  The capital $\Theta$ is used by
convention when the natural parameters are viewed as a matrix, as opposed to
$\theta$ for vector and scalar parameters. 

As Choi matrices are positive semidefinite matrices, there are some known
exponential families parameterized by semidefinite matrices such as the Wishart
distribution \cite{gupta1999matrix,muirhead2009aspects}, or matrix Bingham
\cite{chikuse2012statistics}.  However, neither of these distributions respect
the TP property of Choi matrices.
%
Instead of defining a distribution directly on the space of Choi matrices with
Kraus rank $\leq M$, we consider instead using the statistic defined by mapping
random $MN\times N$ matrix elements $\xi$ with orthonormal columns to the Choi
matrix defined by the relationship in Eq.~\eqref{eq:decomp}.  With some abuse
of notation, let $S(\xi)$ denote the $N^2\times N^2$ matrix defined by
performing the inverse of the column arrangement defined in
Eq.~\eqref{eq:decomp}, and let $\Lambda(\xi)=S(\xi)^\dagger S(\xi)$.  Then, the
exponential family we should consider has the form:
\begin{equation}\begin{aligned}\label{eq:suff2}
    p_\mu(\xi;\Theta)&=\exp\left(\Tr(\Theta^\dagger
    \Lambda(\xi))-\psi(\Theta)+\kappa(\xi)\right)\\ 
    &=\exp\left(\langle\langle
    \xi|(\Theta^\dagger\otimes
    I_M)|\xi\rangle\rangle-\psi(\Theta)+\kappa(\xi)\right)\,.
\end{aligned}\end{equation}
where we have denoted this distribution by $p_\mu$ since it is the
characterized by the average Choi matrix.

Eqs.~\eqref{eq:suff1} and \eqref{eq:suff2} are superficially similar, but it is
important to understand that Eq.~\eqref{eq:suff2} is defined on a completely
different space, and thus the respective normalizers $\psi$ and carrier
measures $\kappa$ are different.  From this point on, we will be considering
distributions defined on Stiefel manifolds, and we will suppress the carrier
measure terms, as it is understood that the distributions are restricted to the
Stiefel manifold of the appropriate dimension.  Since the mapping
$\xi\to\Lambda(\xi)$ is measurable, distributions defined on the Stiefel
manifold generate well defined distributions on the space of Choi matrices.
Furthermore, as the uniform distribution on the Stiefel manifold generates the
distribution defined in Ref.~\cite{bruzda2009random}, we have that the maximum
entropy properties of the exponential families on the Stiefel manifold are in
fact maximizing the entropy relative to the distribution from
Ref.~\cite{bruzda2009random}.

Using Eq.~\eqref{eq:born_stiefel3} and letting $\mathcal{A}_{i,j}$ denote
$(\Theta^\dagger\otimes I_M)_{[i,j]}$ the $(i,j)$th $MN\times MN$ subblock of
$\Theta^\dagger\otimes I_M$ we have
\begin{equation}
    \Tr(\Theta^\dagger\Lambda(\xi))=\sum_{i,j=1}^N\xi_i^\dagger
    \mathcal{A}_{i,j}\xi_j\,.
\end{equation}
%
On a (real-valued) Stiefel manifold, distributions of the form
\begin{equation}\label{eq:gfb}
    p(\xi;\{\mathcal{A}_{i,j}\})=\exp\left(\sum_{i,j=1}^N \xi_i^\dagger
    \mathcal{A}_{i,j}\xi_j -\psi(\{\mathcal{A}_{i,j}\})\right)
\end{equation}
are generalizations of the frame-Bingham distributions
\cite{arnold2013statistics,kume2013saddlepoint}.

Here, we are concerned with a complex-valued manifold, but this is easily
extensible via the standard tricks for converting a complex matrix to a real
one via stacking (see \cite{kent2013new} for an example of this technique as
applied to the traditional Bingham distribution). Strictly speaking, the
structure of $\mathcal{A}_{i,j}$ is more constrained (i.e., each $\mathcal{A}$
is comprised of blocks of scaled identity matrices) than the most general form
of the frame-Bingham distribution, so this is technically a sub-model of the
generalized frame-Bingham distribution.

The frame-Bingham distribution can be Gibbs sampled via the techniques of
\cite{hoff2009simulation} as per the discussion in
\cite{kume2013saddlepoint,arnold2013statistics}, and the generalized case
follows immediately from the scheme described there.  As far as inference
procedures for the frame-Bingham distribution, \cite{kume2013saddlepoint}
introduces a procedure for approximating the normalizer, but we conjecture that
given the additional structure imposed by
$\mathcal{A}_{i,j}$ the estimation process is
replicated using the traditional Bingham distribution, procedures for which can
be found in \cite{chikuse2012statistics}.  Showing this explicitly
is an area of future research.

A closed-form mapping between $\Theta$ and $\Lambda(\xi)$ is not known.
However,
since this is a special case of a generalized frame-Bingham distribution, which
is ultimately derived from a normal distribution using vectorization arguments
\cite{kume2013saddlepoint}, we know some properties of $\Theta$.  First, $\Theta$ is Hermitian and positive
semidefinite. Second, $\Theta$ and $\Lambda(\xi)$ are jointly diagonalizable
(i.e., they have the same eigenvectors), so that the estimation of $\Theta$
from $E[\Lambda(\xi)]$ ultimately amounts to estimating the eigenvalues of
$\Theta$ which are then interpreted as concentration parameters.  Furthermore,
we can assume that the minimum eigenvalue of $\Theta$ is zero.  

\subsection{Binomial Induced Distribution}
%
%
%

Another exponential family can be defined on the space of CPTP maps using the
tomographic experiments $A_i$.
Since the individual $p_i=\Tr(A_i^\dagger \Lambda)$ can be expressed in terms
of Stiefel manifold elements $\xi$ by $p_i=\langle\langle\xi|(A_i\otimes
I_M)|\xi\rangle\rangle$, then $p(\mathbf{x}|\mathbf{n},\Lambda)$ can be
expressed in terms of Stiefel manifold elements to define
$p(\mathbf{x}|\mathbf{n},\xi)$. In turn, given a set of counts $\mathbf{x}$ and
$\mathbf{n}$, the likelihood of $\xi$ given $\mathbf{x}$ and $\mathbf{n}$ can
be used to define a probability distribution on $V_N(\mathbb{C}^{MN})$ by
%
%
\begin{multline}\label{eq:posterior}
    p(\xi|\mathbf{x},\mathbf{n})=\frac{\prod_{i=1}^m
    \binom{n_i}{x_i}p_i^{x_i}(1-p_i)^{n_i-x_i}}
    {\int_{V_N\left(\mathbb{C}^{MN}\right)}
    p(\xi'|\mathbf{x},\mathbf{n})d\xi'}\\
    = \exp\left(\sum_{i=1}^m
    x_i\log(p_i)+(n_i-x_i)\log(1-p_i)\right)C(\mathbf{x},\mathbf{n})
\end{multline}
where $C$ includes both the normalization integral and the terms independent of
$\xi$ from the binomial distributions. When the preparation and measurement
experiments are fixed, Eq.~\eqref{eq:posterior} indicates that the
$p(\xi|\mathbf{x},\mathbf{n})$ is itself an exponential family with parameters
$\theta=\mathbf{x}$ and $\nu=\mathbf{n}-\mathbf{x}$ and sufficient statistics 
\begin{equation}\begin{aligned}
    T_1(\xi)=\begin{bmatrix}\log(\langle\langle \xi|A_1\otimes
        I_M|\xi\rangle\rangle)\\
        \vdots\\ \log(\langle\langle \xi|A_m\otimes
    I_M|\xi\rangle\rangle)\end{bmatrix}
\end{aligned}\end{equation}
and
\begin{equation}\begin{aligned}
    T_2(\xi)=\begin{bmatrix}\log(1-\langle\langle \xi|A_1\otimes
        I_M|\xi\rangle\rangle)\\
        \vdots\\ \log(1-\langle\langle \xi|A_m\otimes
    I_M|\xi\rangle\rangle)\end{bmatrix}\,.
\end{aligned}\end{equation}
Next, substitute the above terms to define another parameterized family on the
Stiefel manifold by
\begin{equation}
    p_c(\xi;\theta_1,\theta_2)=\exp(\theta_1^\top T_1(\xi)+\theta_2^\top
    T_2(\xi)-\psi_c(\theta_1,\theta_2))\,,
\end{equation}
where $\psi_c(\theta_1,\theta_2)=-\exp(C(\theta_1,\theta_1+\theta_2))$. We have
denoted this distribution by $p_c$ since it defines the \textit{conjugate}
prior for a binomial likelihood, as discussed in the next section.

\section{Bayesian Process Tomography}
As an alternative to the maximum likelihood approaches discussed in
Section~\ref{sec:basic_pt}, one can use Bayesian methods.  Unlike maximum
likelihood estimation, which generates a point estimate, Bayesian estimation
produces a posterior distribution $p(\Lambda|\mathbf{x},\mathbf{n})$ from a
prior distribution $p(\Lambda)$ and the experimental results $\mathbf{x}$ (and
parameter $\mathbf{n}$) via Bayes' Rule:
\begin{equation}
    p(\Lambda|\mathbf{x},\mathbf{n})=\frac{p(\mathbf{x}|\Lambda,\mathbf{n})p(\Lambda)}{\int
    p(\mathbf{x}|\Lambda',\mathbf{n})p(\Lambda')\,d\Lambda'}\,,
\end{equation}
where $\Lambda'$ is a dummy variable of integration.  This posterior
distribution can then be used to derive a number of point estimates as well as
credibility intervals.  Why one should prefer Bayesian
methods over frequentist (maximum likelihood) methods is beyond the scope of
this work, but in the case of quantum \textit{state} tomography there is
argument that maximum likelihood is flawed \cite{ferrie2018maximum}.
Furthermore, we will show that we have a prior that is uniform with a certain
choice of parameter, and thus the Bayesian estimates in this case are
equivalent to the frequentist approaches.

In order to perform Bayesian estimation, one needs probability distributions on
the appropriate spaces to define priors.  To be effective, these priors need to
effectively capture belief, likely in terms of average quantities such as mean
and variance.  In this context, exponential families are natural choices since
they are completely characterized by average quantities.  Additionally, to
actually perform the estimation, one needs a closed form expression for the
posterior \textit{or} the ability to sample from it. The priors $p_\mu$ and
$p_c$ and the binomial measurement model induce distributions that can be
efficiently sampled via an extension to the Gibbs sampling technique of
\cite{hoff2009simulation}, as described in the appendix.  Furthermore, this
technique can be adapted to additional families of priors and likelihood
functions.


\subsection{Prior Selection}
In the preceding section, we defined two probability distributions on the space
of CPTP maps using the Stiefel manifold representation.  To reiterate, the
first distribution $p_\mu(\xi;\Theta)$ is the maximum entropy distribution
defined by an average Choi matrix, whereas the second
$p_c(\xi;\theta_1,\theta_2)$ defines the conjugate prior for
$p(\xi|\mathbf{x},\mathbf{n})$.  By this we mean 
\begin{equation}\begin{aligned}
    p(\xi|\mathbf{x},\mathbf{n},\theta_1,\theta_2)&\triangleq
    \frac{p(\mathbf{x}|\mathbf{n},\xi)p(\xi;\theta_1,\theta_2)}{\int_{V_N(\mathbb{C}^{MN})}p(\mathbf{x}|\xi',\mathbf{n})p(\xi';\theta_1,\theta_2)}\\
    &=p(\xi|\mathbf{x}+\theta_1,\mathbf{n}+\theta_1+\theta_2)\,,
\end{aligned}\end{equation}
in other words, using $p_c$ as a prior maintains closure under the
binomial-induced likelihood function and the incorporation of such a prior is
equivalent to having performed a additional experiments that generated a vector
of $\theta_1$ successes in $\theta_1+\theta_2$ trials and adding this to the
original $\mathbf{x}$ and $\mathbf{n}$.

Of course, these priors can be combined and since both are exponential families
$p_{tot}(\xi;\Theta,\theta_1,\theta_2)=p_\mu(\xi;\Theta)p_c(\xi;\theta_1,\theta_2)$
is also an exponential family.
However, since the prior $p_c$ manifests as additional data for the binomially
distributed tomography experiments, we will without loss of generality
primarily focus the analysis on $p_\mu$ (i.e., the impact of the parameters
$\theta_1$ and $\theta_2$ can be equivalently analyzed in the context of
posterior distributions).

Unlike $p_c$, using $p_\mu$ as a prior does not ``fuse'' cleanly with the
binomial experiment data.  However, many quantum characterization experiments
do not report the exact measurement counts used to produce the estimate of
the CPTP map, and thus the prior $p_\mu$ is appropriate for process tomography
when only the map itself is given.  In this case, the choice of $\Theta$ for a
given $\Lambda$ is not known in closed form, but as discussed in
Section~\ref{sec:suff}, $\Lambda$ and $\Theta$ are jointly diagonalizable and
thus share the same eigenvectors.  This leaves the identification of $N^2-1$
eigenvalues of $\Theta$ (since the smallest can be assumed to be 0) which will
be discussed in the context of a numerical sampling scheme in the following
section.

Another, perhaps more relevant instance when this prior is appropriate is when
only a process fidelity is given or implied by an experiment, such as
randomized benchmarking \cite{knill2008randomized} or one of its variants.  In
this case, the implied average (error) map is a uniform depolarizing channel
$\Lambda_{dp}$ with process fidelity $f$ close to one, i.e., for $N=2$ the Choi
matrix of $\Lambda_{dp}$ is
\begin{equation}
\Lambda_{dp} = f|I\rangle\rangle\langle\langle
    I|+\frac{1-f}{3}(|X\rangle\rangle\langle\langle
    X|+|Y\rangle\rangle\langle\langle Y|+|Z\rangle\rangle\langle\langle Z|)\,.
\end{equation}
%
For a multi-qubit system, the eigenvectors of $\Lambda_{dp}$ are the (scaled)
vectorized Pauli matrices and it has two unique eigenvalues, $Nf$ and
$\frac{N}{N^2-1}(1-f)$ (the latter with multiplicity $N^2-1$).  Thus  we have
that $\Theta=\alpha |I_N\rangle\rangle\langle\langle I_N|$ for some positive,
one dimensional $\alpha$.
Also, note that in this case since we are using
$E\left[\Tr\left(|I_N\rangle\rangle\langle\langle I_N|\Lambda\right)\right]$ as
a sufficient statistic, we are effectively defining the exponential family
(i.e., maximum entropy distribution) for which average process fidelity is a
sufficient statistic. 

\begin{figure}[h!] \centering
        \includegraphics[width=1.0\columnwidth]{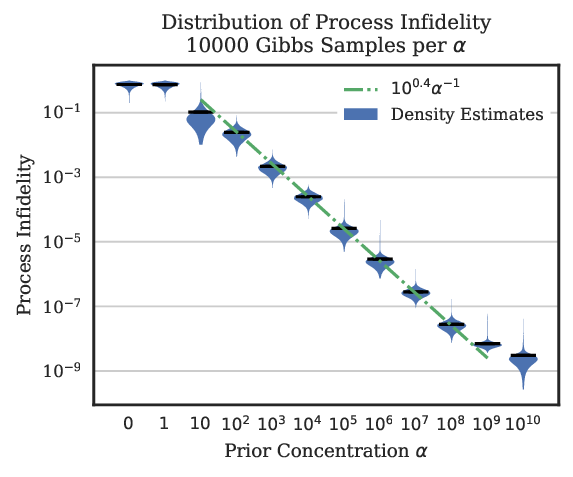}
    \caption{\label{fig:fid_dists} Violin plots showing distributions of
    process infidelity $1-f$ ($M=1000$ samples) drawn from
    $p_\mu(\xi;\alpha|I_2\rangle\rangle\langle\langle I_2|)$ The horizontal
    dashes indicate the means, and a regression fit for the means in the range
    $\alpha\in[10^2,10^8]$ is superimposed.  This illustrates a practical
    relationship between the concentration of the prior $\alpha$ and the
    average process infidelity of resulting random CPTP maps that can be used
    for setting informative priors, from e.g., RB.  Furthermore, we conjecture
    that this trend persists beyond $\alpha=10^8$ if not for numerical
    precision issues.
    } \end{figure}

With regards to selecting a specific value of $\alpha$,
Fig.~\ref{fig:fid_dists} shows distributions of the process \textit{infidelity}
$1-f$ of 1000 randomly generated CPTP maps generated via Gibbs sampling of
$p_\mu\left(\xi;\alpha|I_2\rangle\rangle\langle\langle I_2|\right)$ for a
number of different $\alpha$.  Process infidelity is shown for purposes of
presentation on the logarithmic scale.  The data are displayed using violin
plots, which show both a kernel density estimate of probability distribution of
the data (reflected on both sides of the range bars) as well as the range and
mean \cite{hintze1998violin}. 

Note that the increase in $\alpha$ scales roughly linearly with the process
infidelity in the log-log scale for a wide range of infidelities.  Furthermore,
we suspect that this trend would continue for larger $\alpha$ if not for
numerical issues, as $10^{-8}$ squared is approximately machine precision
for 64 bit floating point numbers.  
Performing a regression on the transformed quantities, we arrive at
$f\approx1-10^{0.4}\alpha^{-1}$ which appears to be accurate for $\alpha\geq100$,
corresponding to a fidelity of roughly $0.97$.
In other words, as the average of the distribution approaches the identity
operation, the largest eigenvalue of the parameter matrix $\Theta$ must
outscale the others and approach infinity, representing infinite concentration
at the identity map.  

The properties of frame-Bingham distribution imply that this eigenvalue scaling
(and its impact on infidelity as a prior for fixed dimension $N$) holds for any
target unitary operation since the particular target unitary only changes the
eigenvectors of the $\Theta$ used, i.e., $\Theta=\alpha|U\rangle\langle U|$ for
for an arbitrary unitary operation $U$.
Furthermore, in the more general case of an average Choi matrix that does not
correspond to a uniform depolarizing channel but is still ``nearly'' a unitary
operation, the observed growth of the dominant eigenvector of $\Theta$ implies
that the additional contributions due to the smaller eigenvalues is negligible. 
Thus, the single parameter prior $\Theta=\alpha|U\rangle\rangle\langle\langle
U|$ will be a good choice for most practical applications.

%
%



\subsection{Bayesian Point Estimation}

The maximum likelihood approach to estimation produces a so-called point
estimate $\hat{\Lambda}_{MLE}$, whereas the Bayesian approach to estimation
produces a probability distribution on $\Lambda$, as discussed above.  Again,
we reiterate that the mapping from $\xi$ to $\Lambda$ is measurable so $p_\mu$
and $p_c$ induce well defined distributions on the space of Choi matrices and
thus we will be somewhat loose in applying these distributions to both cases.
In the context of Bayesian estimation, two common approaches to producing point
estimates are the maximum a posteriori (MAP) and expected a posteriori (EAP)
estimators.

\subsubsection{MAP Estimation Using Exponential Families}
The MAP approach defines a point estimator
\begin{equation}\label{eq:MAPest}
    \hat{\Lambda}_{MAP} = \argmax_{\Lambda\textrm{
        CPTP}}\,\,p(\Lambda|\mathbf{x},\mathbf{n})=\argmax_{\Lambda\textrm{
        CPTP}}\,\,p(\mathbf{x}|\Lambda,\mathbf{n})p(\Lambda),
\end{equation}
since the normalizing term is constant in $\Lambda$ once $\mathbf{x}$ and the
prior distribution have been fixed. Exponential family priors for MAP
estimation have a special relationship with log-likelihood procedures for
maximum likelihood estimation.  Given an arbitrary exponential family prior for
CPTP maps, recall that the functional form of such distribution would be
\begin{equation}
    p_{exp}(\xi|\vartheta) = \exp(\langle \vartheta,T(\xi)\rangle
    -\psi(\vartheta))
\end{equation}
where we have used $\vartheta$ to represent arbitrary
natural parameters corresponding to some arbitrary sufficient statistics
$T(\xi)$.  Furthermore, we have expressed the exponential family as a
distribution on the Stiefel manifold, but in principle exponential families
defined for other representations of CPTP maps would apply to the discussion
below.  Using $p_{exp}$ as a prior results in a posterior distribution
\begin{equation}
    \begin{aligned}
        p(\xi|\mathbf{x},\mathbf{n}) &\propto
        p(\mathbf{x}|\xi,\mathbf{n})p_{exp}(\xi|\vartheta)\\
    %
\end{aligned}
\end{equation}
which is again an exponential family.

From a MAP estimation perspective, 
\begin{equation}
    \begin{aligned}
        \hat{\Lambda}_{MAP} & = \Lambda \left(\argmax_{\mathcal{\xi}}\,\,
        \log\left(p(\mathbf{x}|\xi,\mathbf{n})p_{exp}(\xi|\vartheta)\right)\right)\\
        & =
        \Lambda\left(\argmax_{\xi}\,\,\log\left(p(\mathbf{x}|\xi,\mathbf{n})\right)+\langle\vartheta,T(\xi)\rangle\right).
    \end{aligned}
\end{equation}
Thus, the use of an exponential family prior can be applied to a log-likelihood
based maximum likelihood estimation routine by adding the term
$\langle(\vartheta,T(\xi)\rangle$.  In particular, for our combined prior
$p_{tot}(\xi;\Theta,\theta_1,\theta_2)$  we have
\begin{equation}
    \begin{aligned}
        \hat{\Lambda}_{MAP}=&\Lambda\bigg(\arg\max_{\xi}
        \,\,\log\left(p(\mathbf{x}+\theta_1|\xi,\mathbf{n}+\theta_1+\theta_2)\right)\\
        &+\Tr\left(\Theta^\dagger\Lambda(\xi)\right)\bigg)
    \end{aligned}
\end{equation}
where setting $\Theta=0$ defines the uniform distribution on the Stiefel
manifold, and the MAP estimate reduces to MLE estimate.

Since the prior $p_c$ folds in the binomial data in the likelihood this prior
requires no changes to an existing maximum likelihood estimator.  Indeed, this
is equivalent to using conjugate priors on each tomographic experiment
individually.  The prior $p_\mu$ incorporates a single term in the
log-likelihood estimator that is straight-forward to implement but obviously needs to
be modified depending on the matrix representation used for the CPTP maps. This
applies to Liouvillian superoperators through $\mathfrak{P}(\Theta)$ and
for other linear transformations in an analogous manner.  We have made these
modifications to the gradient-based approach of \cite{knee2018maximum} which
affects both the objective function, but also the rules for step size.
Simulated results are shown in Section~\ref{sec:sim_point}.

\subsubsection{EAP Estimation}

The EAP estimate $\hat{\Lambda}_{EAP}$ is simply the mean of the posterior
distribution, which to under an arbitrary prior with parameter $\vartheta$ is
expressed as
\begin{equation} \hat{\Lambda}_{EAP} =
    \int_{V_N(\mathbb{C}^{MN})}\Lambda(\xi)p(\xi|\mathbf{x},\mathbf{n},\vartheta)
\end{equation}
This also has the interpretation of minimizing the mean squared error with
respect to the posterior distribution. In other words, EAP estimates minimize
the expected value of the loss function
$L(\hat{\Lambda},\Lambda)=||\hat{\Lambda}-\Lambda||_2^2$.  Other loss functions
(i.e., different Bayes risks) can be used in an analogous fashion.

From a practical perspective, since the normalization constants in the
posterior distribution for either prior $p_\mu$ and $p_c$ are intractable, this
average is computed numerically using the Gibbs sampling routine.  
A sufficient number of samples from the posterior distribution
$\{\xi^{(i)}\}\sim p(\xi|\mathbf{x},\mathbf{n},\Theta,\theta_1,\theta_2)$ are
used to produce to produce an accurate average
$\hat{\Lambda}=\frac{1}{|\{\xi^{(i)}\}|}\sum_i\Lambda(\xi^{(i)})$.
%
Sampling from the posterior is more computationally intensive than the MLE or
MAP approaches, and in our experience can have slow convergence  and the
potential for highly correlated samples, both of which require additional
samples to be produced to maintain an effective sample size for an accurate
estimate.  That said, the production of EAP estimates generate approximations
of the full posterior distribution allowing for the generation of credibility
intervals as well as posterior distributions of other statistics, such as
diamond norm (see Section~\ref{sec:full_bayes}).

\subsubsection{Simulation of Point Estimators}\label{sec:sim_point}

Simulated process tomography was performed using perfect state preparation and
measurement for all combinations of the states $|0\rangle$, $|1\rangle$,
$|+\rangle$ and $|-\rangle$.
Ideal state preparation and measurement (SPAM) was used to clearly illustrate
the impact of the prior on the estimation process without confounding factors.
As maximum likelihood process tomography is a special case of the Bayesian
approach presented here, the presence of SPAM errors should impact this
Bayesian approach in much the same way as standard process tomography (see
e.g., \cite{Merkel2013} for a brief discussion), but incorporation of prior
information from SPAM agnostic procedures (such as RB) should bias the results
towards the true fidelity.  Incorporation of these priors into a
self-consistent form of process tomography (such as GST
\cite{Merkel2013,blume2016certifying}) would address any concerns with SPAM
errors and should be easily implemented for MAP estimates (see also
Section~\ref{sec:conc} for additional discussion), and we leave this for future
work.
Each of the sixteen input/output combinations were repeated $n$ times varying
over the set $\{10,10^2,10^3,10^4, 10^5\}$.
Two sets of 100 ground truth CPTP maps were generated from
$p_\mu(\xi;\alpha|I\rangle\rangle\langle\langle I|)$ with $\alpha=10^2$ and
$10^4$, corresponding to average process fidelities of approximately $0.97$ and
$0.9997$, respectively.  These two priors, as well as the uniform $\alpha=0$
prior were used to perform MAP and EAP estimation.  For the MAP estimates we
used a slightly modified version of \cite{knee2018maximum}, and note that the
$\alpha=0$ prior corresponds to MLE.  For the EAP estimates, 1000 burned in
Gibbs samples from the sampler were used.

\begin{figure}[h!] \centering
        \includegraphics[width=1.0\columnwidth]{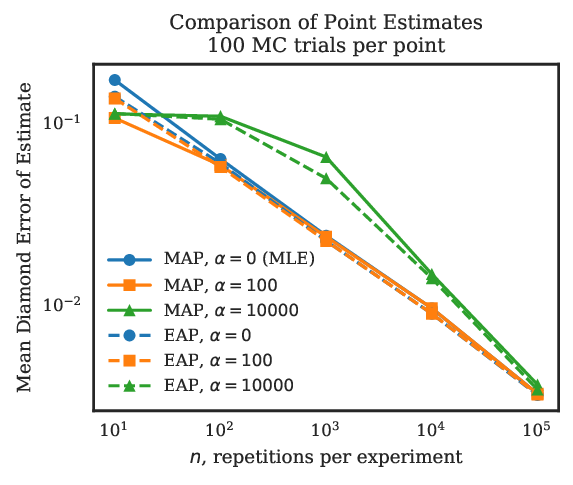}
    \caption{\label{fig:diam_100} 
    Comparison of mean diamond error to
    ground truth for
    different combinations of estimators (MLE, MAP, and EAP) and priors (defined by parameter
    $\alpha|I\rangle\rangle\langle\langle I|$).  
    Ground truth
    gates were drawn from the distribution
    $p_\mu(\xi;\alpha|I\rangle\rangle\langle\langle I|)$ with $\alpha=100$,
    which corresponds to an average process fidelity of $\approx0.97$.
    The
    priors correspond to average process fidelities of $0.25$, $\approx0.97$
    and $\approx0.9997$ for $\alpha=0$, $100$, and $10000$, respectively.
    The Bayesian estimates with the correct ($\alpha=100$) prior produce slight
    improvements over MLE but due to the relative weakness of the prior this
    advantage is quickly erased with increasing $n$.  The prior with
    $\alpha=10000$ is highly concentrated and it requires considerable data to
    overcome this mismatched prior, as the ground truth samples are highly
    unlikely for this prior.
    } 

\end{figure}

\begin{figure}[h!] \centering
        \includegraphics[width=1.0\columnwidth]{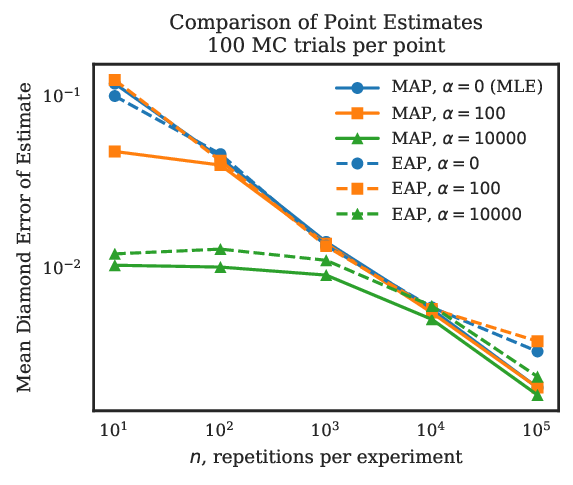}
    \caption{\label{fig:diam_10000} 
    Comparison of mean diamond error to ground truth for the same combination
    of estimators and priors as Fig.~\ref{fig:diam_100}.
    Here, however, ground truth gates were drawn from the prior
    $p_\mu(\xi;\alpha|I\rangle\rangle\langle\langle I|)$ with $\alpha=10000$,
    which corresponds to an average process fidelity of $\approx0.9997$.
    Given the high concentration of the ground truth distribution the
    advantages of the matched prior are shown with substantial improvement in
    estimation error for small $n$.
    We believe the divergence in the EAP estimators for $n=10^5$ is likely due
    to occasional poor convergence in the underlying MCMC samples which can be
    avoided via manual inspection (to continue to produce samples) but is
    difficult to avoid in an automated fashion.
    %
    } 

\end{figure}

Figures~\ref{fig:diam_100}~and~\ref{fig:diam_10000} show the mean diamond norm
error between the various point estimates and the two sets of ground truth maps
used above.  This is analogous to the root-mean-square error of the estimator.
In accordance with general estimation theory, the estimators tend to converge
to the true value as the number of measurements increases, and this this rate
of decrease is eventually roughly proportional to the square root of the number
of measurements.  These figures also demonstrate how the priors interact with
the data, in that an informative prior can bias the estimate and result in a
more accurate estimate given fewer measurements (Fig.~\ref{fig:diam_10000}),
but can also bias the estimate in a negative manner when the true map is
unlikely given a highly concentrated prior (Fig.~\ref{fig:diam_100}).
Additionally, it appears that the diffuse prior cases in
Fig.~\ref{fig:diam_10000} are suffering reduced effective sample size, since
the EAP estimates diverge slightly from the MAP estimates as the sample sizes
increase.  This issue is is really only a problem in the automated analysis
here, for analysis of a single experiment, visual inspection of the samples or
other methods for estimating effective sample size could be employed.

\begin{figure}[h!] \centering
        \includegraphics[width=1.0\columnwidth]{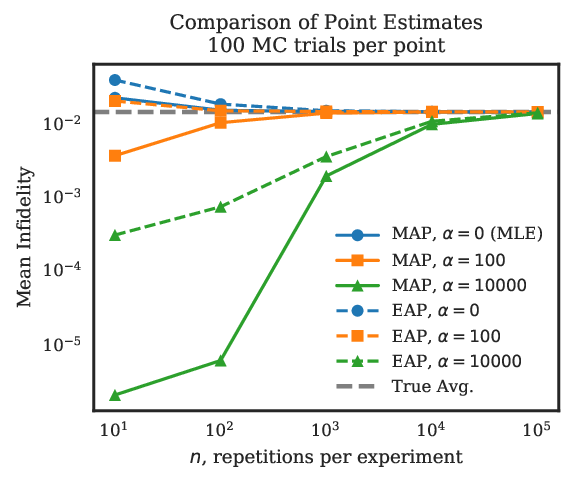}
    \caption{\label{fig:inf_100} Comparison of mean process infidelity to the
    identity map for the same combinations of estimators and priors as
    Fig.~\ref{fig:diam_100}, as well as the sample average of the process
    infidelities of the ground truth gates.  Ground truth gates were drawn from
    the prior $p_\mu(\xi;\alpha|I\rangle\rangle\langle\langle I|)$ with
    $\alpha=100$ corresponding to an average process fidelity of $\approx0.97$.  
    As expected, the MAP estimates are highly biased with respect to the
    process infidelity, as compared with the EAP estimates, and the EAP
    estimates with $\alpha=100$ produce the closest posterior infidelities to
    the truth data.  Note, however,
    that despite large deviations in the process infidelity of the estimates,
    the diamond errors between the estimates and true gates are similar (see
    Fig.~\ref{fig:diam_100}).
    } 

\end{figure}

\begin{figure}[h!] \centering
        \includegraphics[width=1.0\columnwidth]{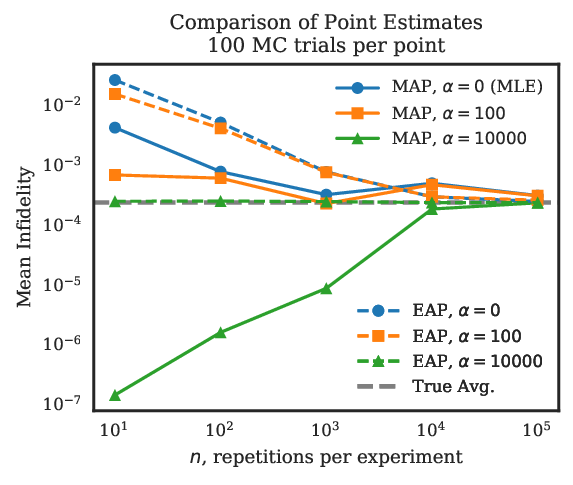}
    \caption{\label{fig:inf_10000} Comparison of mean process infidelity to the
    identity map for the same combinations of estimators and priors as
    Fig.~\ref{fig:diam_100}, as well as the sample average of the process
    infidelities of the ground truth gates.  Ground truth
    gates were drawn from the prior
    $p_\mu(\xi;\alpha|I\rangle\rangle\langle\langle I|)$ with $\alpha=10000$
    corresponding to an average process fidelity of approximately $0.9997$.
    The trends here are essentially the same as Fig.~\ref{fig:inf_100}, but
    even more pronounced.
    } 

\end{figure}

Figures~\ref{fig:diam_100} and~\ref{fig:diam_10000} focused on the overall
estimation error between the tomographic reconstruction and the true map.
Since the prior $p_\mu$ is specified in terms of average process fidelity,
analyzing the behavior of the estimators with respect to process fidelity
offers an alternative view to the properties of the estimators.
Figures~\ref{fig:inf_100} and~\ref{fig:inf_10000} show the mean process
infidelity of the estimators as a function of $n$.  As we expect from the
Figs.~\ref{fig:diam_100} and~\ref{fig:diam_10000} the process infidelities
converge across all estimators as $n$ increases.
These figures indicate that MAP and MLE estimation will produce higher
estimates of process fidelity than EAP estimation using the uniformly
depolarizing prior considered here.  In particular, EAP estimates with a
properly matched prior produce the most accurate estimates of process
infidelity.
This discrepancy between the estimated infidelity despite the similarity in
overall diamond error indicates that the MAP (and MLE) approaches are heavily
biased towards the dominant eigenvectors of the prior parameter as compared
with EAP estimation.

\subsection{Posterior Distributions and Credibility
Intervals}\label{sec:full_bayes}

\begin{figure}[h!] \centering
        \includegraphics[width=1.0\columnwidth]{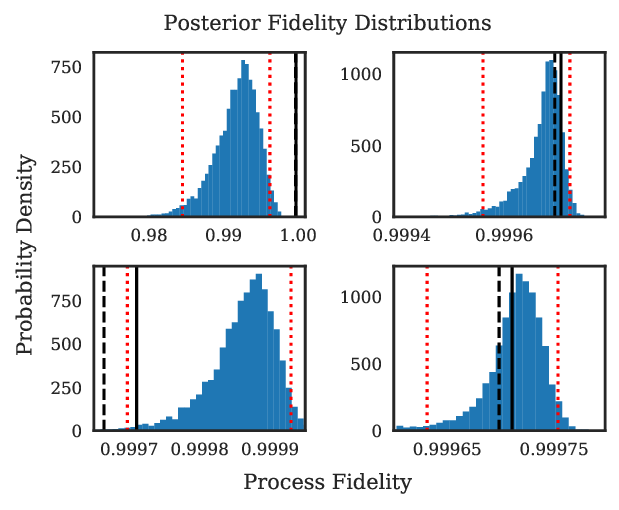}
    \caption{\label{fig:posterior_fidelity} Comparison of posterior
    distributions.  Each subfigure corresponds to 10000 samples from the
    posterior distribution.  Top row: Uniform priors.  Bottom row: prior
    $p_\mu(\xi;10^4|I\rangle\rangle\langle\langle I|)$.  Left column $n_i=100$
    measurements per simulated experiment. Right column $n_i=10^5$ measurements
    per simulated experiment. Black solid line indicates the true process
    fidelity, the dashed black line the fidelity of the MLE.  Red dotted lines denote
    the inner 95\% credibility interval determined by the posterior samples.
    } 

\end{figure}

In addition to point estimation, the Gibbs sampling routine from the posterior
distribution can be used for additional Bayesian approaches such as the
construction of credibility intervals. As an example,
Fig.~\ref{fig:posterior_fidelity} shows histograms of posterior distributions
for different combinations of prior and $\mathbf{n}$, along with the inner 95
percentile of the posterior distribution, i.e., the 95\% credibility interval.
The two upper left panel shows the uniform prior with $n_i=100$ measurements
per state preparation and measurement configuration, the upper right the
uniform prior with $n_i=10^5$, the lower left the prior
$p_\mu(\xi;10^4|I\rangle\rangle\langle\langle I|)$ with $n_i=100$, and the bottom
left panel the prior $p_\mu(\xi;10^4|I\rangle\rangle\langle\langle I|)$ and
$n_i=10^5$.

\begin{figure}[h!] \centering
    \includegraphics[width=1.0\columnwidth]{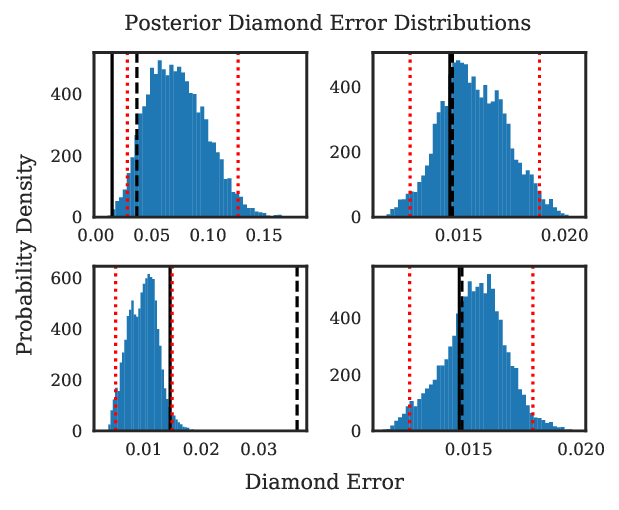}
    \caption{\label{fig:posterior_diamond} Comparison of posterior
    distributions of diamond error.  Each subfigure corresponds to 10000 samples from the
    posterior distribution (the same samples as
    Fig.~\ref{fig:posterior_diamond}).  Top row: Uniform priors.  Bottom row: prior
    $p_\mu(\xi;10^4|I\rangle\rangle\langle\langle I|)$.  Left column $n_i=100$
    measurements per simulated experiment. Right column $n_i=10^5$ measurements
    per simulated experiment. Black solid line indicates the true process
    fidelity, the dashed black line the fidelity of the MLE.  Red dotted lines denote
    the inner 95\% credibility interval determined by the posterior samples.
    These plots highlight the power of a posterior distribution, as we can
    produce estimates and credibility intervals of \textit{functions} of the
    underlying quantum process that are otherwise experimentally inaccessible.
    } 

\end{figure}

Samples from the posterior distributions can also be used to perform Bayesian
analysis of arbitrary functions of the posterior distribution of CPTP maps.  As
an example, Fig.~\ref{fig:posterior_diamond} shows posterior distributions of
the diamond error (from the identity map) for the same data shown in
Fig.~\ref{fig:posterior_fidelity}.  This shouldn't be confused with data
shown in Figs.~\ref{fig:diam_100} and~\ref{fig:diam_10000} which show the
diamond error between the point estimates and the true gates across an
ensemble of simulated tomography experiments.  The plots on the
left illustrate the strength of the prior in terms of its concentration in
diamond norm.  The uniform prior for the $n_i=100$ case is considerably less
concentrated than for the $\alpha=10^4$ prior.  Additionally, since the true
map was drawn from the $\alpha=10^4$ prior, the true diamond error is in the
95\% posterior credibility interval for the bottom-left plot, but not when the
uniform prior is used.  However, when $n_i$ is large (corresponding to the
right column) the posterior distributions (and resulting credibility intervals)
are much closer.

\section{Conclusion}\label{sec:conc}
In this work we have used the theory of exponential families of probability
distributions using Stiefel manifolds as a sample space to induce distributions
on the space of CPTP maps.  These distributions are used in Bayesian analysis
of process tomography as both prior and posterior distributions.  From the
perspective of priors, one of these distributions is the maximum entropy
distribution defined by an average Choi matrix, whereas the other is equivalent
to the conjugate prior for the binomial distributions that underlie process
tomography experiments.  We compared Bayesian MAP and EAP point estimators and
discussed some impacts of the priors on the estimators.  Additionally, we
showed that the Gibbs sampling approach can be used to produce posterior
credibility intervals for parameters of interest.

The analysis and examples presented in this manuscript are by no means
exhaustive, the primary focus was on demonstrating that the priors and sampling
approaches work correctly, and that the results agree with the general theory
of Bayesian estimation in a classical context.  In particular, for the full
Bayesian analysis, we only considered some basic parameters of CPTP maps, one
could envision other parameters of interest such as nonunitarity (i.e.,
translation of the center of the Bloch sphere via the CPTP map).  Another
aspect of Bayesian estimation which we have not touched are sequential
techniques.  The distributions discussed here could be used as proposal
distributions for the techniques proposed in
\cite{granade2016practical,pogorelov2017experimental}.

As far as future extensions, we note that we have already shown how the
method of \cite{hoff2009simulation} can be extended to include essentially
arbitrary functions of a CPTP map (here we used $\log$).  This observation
allows for the definition and sampling of a wide range of distributions on the
space of CPTP maps.  
For example, the addition of regularizers or penalty terms to a maximum
likelihood estimation process can often be interpreted in Bayesian context as
the component due to a particular choice of (exponential family) prior in a MAP
process \cite{seeger2010variational,stuart2010inverse}. Thus sparsity-enforcing
regularization terms used for process tomography (e.g., an $\ell_1$ penalizer
as in \cite{mohseni2008quantum}) can be interpreted as MAP estimates using an
exponential family. Considering $||\Lambda||^2_2$ or $||\Lambda||_1$ as
parameters in an exponential family (perhaps expressed in a Pauli Basis) appear
to be implementable in the Gibbs sampling framework, and thus the use of these
sorts of regularization terms can be analyzed in the fully Bayesian context.
Many other natural quantities of interest can be expressed as simple functions
of $\Lambda$, and thus can be interpreted in the framework of exponential
families on Stiefel manifolds.

The other major extension that we see for these concepts is to define joint
distributions between CPTP maps.  In one sense, since $\exp$ and the Gibbs
sampling framework naturally factorizes, the algorithm can be extended to add
another layer of conditional sampling.  However, for more complex tomographic
procedures such as gate set tomography \cite{Merkel2013,blume2016certifying},
the likelihood terms will be higher order polynomials in the random gates and
will likely require clever handling in the inner loops of the sampling routine
to remain computationally tractable.

Aside from the applications of these distributions to process
tomography we note that these distributions could also be applied to circuit
simulation to study the effects of non-Pauli errors in circuit simulation and
threshold computations.  
Related to this, we note that a future direction of research is the
distribution on quantum states induced by the application of these random CPTP
maps to a given input state.
%
Furthermore, it should be possible to introduce correlations between errors and
use a similar Gibbs sampling approach to generate random errors that are
correlated in both time and space.

\acknowledgments{
We thank Dave Clader, Gregory Quiroz, and Dennis Lucarelli for their useful
comments in the preparation of this manuscript. This project was supported by
the Intelligence Advanced Research Projects Activity via Department of Interior
National Business Center contract number 2012-12050800010. The U.S. Government
is authorized to reproduce and distribute reprints for Governmental purposes
notwithstanding any copyright annotation thereon. The views and conclusions
contained herein are those of the authors and should not be interpreted as
necessarily representing the official policies or endorsements, either
expressed or implied, of IARPA, DoI/NBC, or the U.S.  Government.
}

\bibliographystyle{apsrev4-1}
\bibliography{references}

\appendix

\section{Sampling Algorithm}

The algorithm of \cite{hoff2009simulation} uses several stages of Markov Chain
Monte Carlo (MCMC) sampling to generate random samples whose stationary
distribution converges to the target distribution.  As noted in the main text,
the algorithm is designed for real-valued Stiefel manifolds, so we use the
procedure in \cite{kent2013new} to convert between $V_N(\mathbb{C}^{MN})$ to
$V_N(\mathbb{R}^{2MN})$. The outermost layer of
sampling produces a new sample $\xi'$ from the current sample $\xi$. Inside
this layer, the next layer selects a column $r$ and then $\xi_r$ is generated
conditionally with the remaining columns (denoted $\xi_{-r}$).  This process is
repeated for a random permutation of $\{1,\dots,N\}$ so that all columns are
sampled, resulting in a new sample $\xi'$.  To sample a column $\xi_r$
conditioned on $\xi_{-r}$, let $z=\mathcal{N}^\dagger\xi_r$, with
$\mathcal{N}$ and orthonormal basis for the null space of $\xi_{-r}$. The final
MCMC stage then samples the coordinates $z_i$ in random order conditioned on the
remaining parameters, denoted $z_{-i}$.  These steps are summarized in
Algorithm~\ref{alg:outerMCMC}, with the further detail and derivation of the
underlying distributions following.

\begin{figure}[h!]
\begin{algorithm}[H]
    \caption{Sampling $p(\xi'|\xi)$}
    \label{alg:outerMCMC}
    \begin{algorithmic}[l]
        \Function{OuterMCMCStep}{$\xi,\Theta,\{A_j\}, \mathbf{x},\mathbf{n}$}
        \For{$r$ in \texttt{randperm}$(\{1,\dots,N\})$}
        \State $\mathcal{N}\gets I-\xi_r \xi_r^\dagger$
        \State $z\gets\mathcal{N}^\dagger\xi_r$
        \For{$i$ in \texttt{randperm}$(\{1,\dots,2MN-1\}$}
        \State $\tau\gets z_i^2$
        \State $s\gets\texttt{sign}(z)$
        \State $q_{-i}\gets z_{-i}^2/(1-\tau)$
        \State Sample $\tau'\propto
        p(\tau',1|q_{-i},s_{-i})+p(t,-1|q_{-i},s_{-i})$ 
        \State Sample $s'\propto
        p(\tau',s'|q_{-i},s_{-i})+p(\tau',s'|q_{-i},s_{-i})$
        \State $z\gets \frac{\sqrt{1-\tau'}}{\sqrt{1-\tau}} z$
        \State $z_i\gets \sqrt{\tau'}s$
        \EndFor
        \State $\xi_r\gets \mathcal{N}z$
    \EndFor

    \State\Return $\xi$
    \EndFunction

\end{algorithmic}
\end{algorithm}
\end{figure}


From the above algorithm, one can see that the true difficulty lies in sampling
$\tau'$ and $s'$, with the remainder of the algorithm being an exercise in
linear algebra.  The conditional posterior distribution of the column $\xi_r$ (without loss of
generality we ignore $\theta_1$ and $\theta_2$) is given by
\begin{multline}
    \label{eq:A1}
    p(\xi_r|\xi_{-r},\Theta,\mathbf{x},\mathbf{n}) \propto
    \exp\Biggl(\xi_r^\dagger(\Theta\otimes I_M)_{[r,r]}\xi_r\\
    +2\Re\biggl(\sum_{j\neq r}\xi_{j}^\dagger(\Theta\otimes I_M)_{[j,r]}\xi_r\biggr)\\
    +\sum_{j=1}^m x_j\log\Biggl[\xi_r^\dagger(A_j\otimes I_M)_{[r,r]}\xi_r\\
    +2\Re\biggl(\sum_{k\neq r}\xi_k^\dagger(A_j\otimes
    I_M)_{[k,r]}\xi_r\biggr)+\sum_{k\neq r}\xi_k^\dagger (A_j\otimes
    I_M)_{[k,k]}\xi_k\Biggr]\\
    +\sum_{j=1}^m(n_j-x_j)\log\Biggl[1-\xi_r^\dagger(A_j\otimes
    I_M)_{[r,r]}\xi_r\\
    -2\Re\biggl(\sum_{k\neq r}\xi_k^\dagger(A_j\otimes
    I_M)_{[k,r]}\xi_r\biggr)-\sum_{k\neq r}\xi_k^\dagger (A_j\otimes
    I_M)_{[k,k]}\xi_k\Biggr]\Biggr)\,.\\
\end{multline}
Applying the substitution $z=\mathcal{N}^\dagger \xi_r$ and gathering like
terms, the conditional posterior distribution of an element $z_i$ given the
remaining elements $z_{-i}$ (and $\xi_{-r}$, etc.,) is 
\begin{multline}
    \label{eq:A2}
    p(z_i|z_{i-1},\dots)\propto \exp\biggl(bz_i^2+z_{-i}^\dagger
    Bz_i+2\Re(c_i^\dagger z_i+c_{-i}^\dagger z_{-i})\\
    +\sum_{j=1}^m x_j
    \log\biggl[\tilde{b}_jz_{i}^2+z_{-i}^\dagger\tilde{B}_jz_{-i}\\
    +2\Re((\tilde{c}_j)^\dagger_iz_i+(\tilde{c}_j)^\dagger_{-i}z_{-i})+\tilde{d}_j\biggr]\\
    +\sum_{j=1}^m (n_j-x_j)
    \log\biggl[1-\tilde{b}_jz_{i}^2-z_{-i}^\dagger\tilde{B}_jz_{-i}\\
    -2\Re((\tilde{c}_j)^\dagger_iz_i+(\tilde{c}_j)^\dagger_{-i}z_{-i})-\tilde{d}_j)\biggr]\biggr),\\
\end{multline}
where the terms $b$, $\tilde{b}_j$, and $d_j$ are scalars, $c$ and $\tilde{c}_j$ are
vectors, and $B$ and $\tilde{B}_j$ are matrices whose straightforward
computation from $\mathcal{N}$, $\Theta$, $A_i$, and $\xi_j$ we have omitted.
%
%

As in \cite{hoff2009simulation} we perform yet another coordinate
transformation, letting $\tau=z_i^2$, $s_i=\sign(z_i)$,
$q_{-i}=z^2_{-i}/(1-\tau)$ where the exponent here is elementwise, and
$s_{-i}=\sign(z_{-i})$, a vector of signs.  Thus, $z_i=\sqrt{\tau}s_i$ and
$z_{-i}=\sqrt{1-\tau}\sqrt{q_{-i}}\circ s_{-i}$ where $\sqrt{\cdot}$ is
elementwise and $\circ$ denotes elementwise multiplication. 
Applying the chain rule as in \cite[Sec.~3.1]{hoff2009simulation} yields an
expression for the joint distribution of $\tau$ and $s_i$ as
\begin{multline}
    p(\tau,s_i|q_{-i},s_{-i},\dots)\propto\\
    p(z_i=\sqrt{\tau}s_i|z_{-i}=\sqrt{1-\tau}q_{-i}\circ s_{-i},\dots)\\
    \times\tau^{-\frac{1}{2}}(1-\tau)^{\frac{(NM-4)}{2}}\,.
\end{multline}
Sampling $z_i$ then amounts to sampling $\tau$ from
\begin{equation}\label{eq:ptau}
    p(\tau,s_i=1|q_{-i},s_{-i})+p(\tau,s_i=-1|q_{-i},s_{-i})
\end{equation} 
and then sampling $s_i$ given the chosen $\tau$. We found that the
distributions of interest here are often too strongly concentrated to use the
rejection sampling approach in \cite{hoff2009simulation}. Instead, we adopt the
other suggestion in \cite{hoff2009simulation}, using a grid-based sampling of
$\tau$.  The gist of the approach is to start with an initial sample of $K$
evenly spaced $\tau_j$ from $(0,1)$, evaluate their probabilities, and sample
nearby points whose probabilities are within some threshold of the maximum
sampled probability.  This process is repeated, retaining only the likely
samples until an effective sample size criteria is reached.  Here we used a
standard criterion from the field of sequential Monte Carlo
\cite{doucetsequential}, and set the effective sample size to $K_{eff}=(\sum
w_j^2)^{-1}$, where $w_j$ are the normalized weights of the samples $\tau$.
Here, the adaptive sampling procedure was terminated when $K_{eff}\geq
K_{min}=K/10$.  This process is described in Algorithm~\ref{alg:sampletau}.
Once the sampling of $\tau$ has been completed, sampling of $s_i$ is done using
the distribution defined by $p(s_i|\tau,\dot)$ for $s_i\pm1$. Cycling through
Algorithm~\ref{alg:outerMCMC} for all columns will then result in a new sample
for $\xi$.

\begin{figure}[h!]
    \begin{algorithm}[H]
    \caption{Adaptive Grid Sampling of $\tau$}
    \label{alg:sampletau}
    \begin{algorithmic}[l]
        \State $K_{eff}\gets K_{min}-1$
        \State $T\gets$ \texttt{linspace}$(0,1,K)$
        \State $\Delta\gets 1/K$
        \For{$\tau_j\in T$}
        \State  $p_j \gets p(\tau_j,s_i=1|q_{-i},s_{-i})+p(\tau_j,s_i=-1|q_{-i},s_{-i})$
        \State $w_j \gets p_j/\sum p_j$
        \EndFor
        \State $K_{eff}=(\sum p_j^2)^{-1}$
        \While{$K_{eff}<K_{min}$}
        \State $\Delta\gets\Delta/K$
        \State $T'\gets\{\}$
        \For{$\tau_j \in T$}
        \If{$w_j>\varepsilon \max_j{w_j}$}
        \State$T'=T'\cup$\texttt{linspace}$(\tau_i-\frac{\Delta}{2},\tau_j+\frac{\Delta}{2},K)$
        \EndIf
        \EndFor
        \State $T\gets T'$
        \For{$\tau_j\in T$}
        \State  $p_j \gets p(\tau_j,s_i=1|q_{-i},s_{-i})+p(\tau_j,s_i=-1|q_{-i},s_{-i})$
        \State $w_j \gets p_j/\sum p_j$
        \EndFor
        \State $K_{eff}=(\sum p_j^2)^{-1}$
        \EndWhile
        \State Sample $\tau$ from $p(\tau_j)=w_j$
    \end{algorithmic}
\end{algorithm}
\end{figure}

In the simple tomography experiments considered here, the terms in
Eq.~\eqref{eq:ptau} due to the binomial distribution can be computed rapidly
for a set of candidate $\tau_j$ using the vectorization and broadcasting
capabilities of modern computer linear algebra package. This is possible
because the terms in Eq.~\ref{eq:A2} can be expressed in relatively simple
terms once the portions dependent on the other columns $\xi_{-r}$ are fixed. In
a more complex tomographic experiment such as gate set tomography, the
likelihood expressions of the resulting binomial distributions will depend on
several elements from Stiefel manifolds (one for each gate), and while the
joint family will still be an exponential family, the expressions for the
likelihood will likely be more complicated and considerable thought will be
required to make sampling from the grid efficient.

\end{document}